\documentclass[reprint,aps,prl,secnumarabic,balancelastpage,amsmath,amssymb,floatfix,superscriptaddress,nofootinbib]{revtex4-1}
\usepackage[utf8]{inputenc}
\usepackage{amsmath}
\usepackage{amsfonts}
\usepackage{bbold}
\usepackage{amssymb}
\usepackage[pdftex]{graphicx}

\usepackage{hyperref}
\hypersetup{
	colorlinks = true,
	citecolor = {red}
	}

\let\vec\mathbf

\begin{document}

\title{\bf On two phases inside the Bose condensation dome of Yb$_2$Si$_2$O$_7$}

\author{Michael O. Flynn}
\email{miflynn@ucdavis.edu}
\affiliation{Department of Physics, University of California Davis, CA, 95616, USA}
\author{Thomas E. Baker}
%\email{thomas.baker@usherbrooke.ca}
\affiliation{Institut quantique \& D\'epartement de physique, Universit\'e de Sherbrooke, Sherbrooke, Qu\'ebec J1K 2R1 Canada}
\author{Siddharth Jindal}
%\email{sjindal2@illinois.edu}
\affiliation{Department of Physics, University of Illinois Urbana-Champaign}
\author{Rajiv R. P. Singh}
%\email{singh@physics.ucdavis.edu}
\affiliation{Department of Physics, University of California Davis, CA, 95616, USA}

\date{\today}

\begin{abstract}

Recent experimental data on Bose-Einstein Condensation (BEC) of magnons in the spin-gap compound $\text{Yb}_{2}\text{Si}_{2}\text{O}_{7}$ revealed an asymmetric BEC dome \cite{PhysRevLett.123.027201}. We examine modifications to the Heisenberg model on a breathing honeycomb lattice, showing that this physics can be explained by \itshape competing forms of weak anisotropy. \normalfont We employ a gamut of analytical and numerical techniques to show that the anisotropy yields a field driven phase transition from a state with broken Ising symmetry to a phase which breaks no symmetries and crosses over to the polarized limit.

\end{abstract}

\maketitle

In recent decades, models of localized spins have been shown to contain a wealth of familiar and exotic phases of matter. Interesting orders can be achieved by considering models with competing interactions, which naively require the satisfaction of incompatible constraints to achieve a ground state. Nature's creative mechanisms for resolving these tensions within quantum mechanics is responsible for much of the diversity of phenomena observed within many-body theory \cite{Savary_2016,Gingras_2014,Li_2016,Balents2010SpinLI,RevModPhys.51.659,RevModPhys.91.041003,fradkin_2013}.

\begin{figure}
  \includegraphics[width=0.75\linewidth]{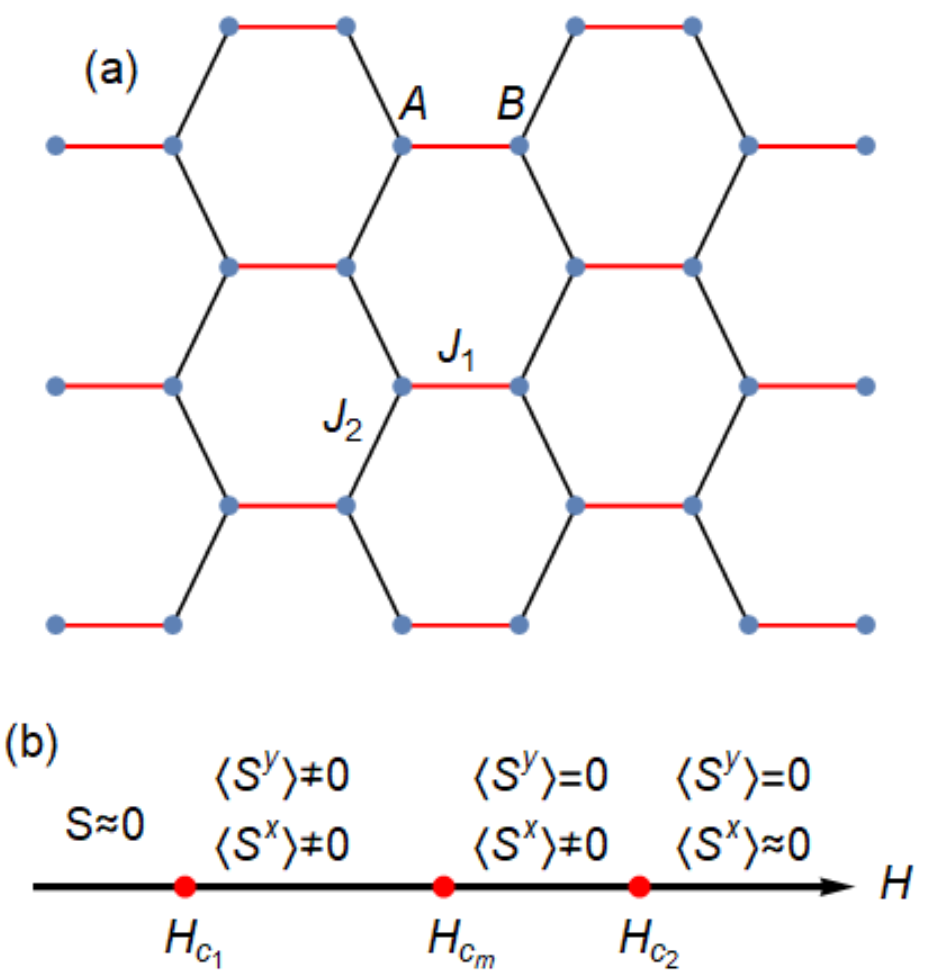}
  \caption{(a) A section of the honeycomb lattice. Each spin (blue dots) has a preferred neighbor (red bonds) which it interacts with more strongly than others: $J_{1}>J_{2}$. The ground state in zero field is a product of spin singlets along the red bonds. (b) Schematic $T=0$ phase diagram obtained from DMRG and mean-field theory. From left to right, the phases are a global spin singlet, $\mathbb{Z}_{2}$ symmetry breaking antiferromagnet, canted antiferromagnet, and the polarized phase. The critical points $H_{c_{1}}$ and $H_{c_{m}}$ are in the Ising universality class while $H_{c_{2}}$ is a crossover.}
  \label{fig:lattice}
\end{figure}

A clear example of such physics is found in dimer magnetism, where antiferromagnetic behavior is brought into tension with polarizing magnetic fields \cite{PhysRevLett.123.027201,RevModPhys.86.563,PhysRevLett.96.077204,Kofu_2009,TSUI20071319,article}. In these systems, spins tend to pair into singlets in the low-field ground state. A simple example of this phenomenon is realized in the antiferromagnetic Heisenberg model on the breathing honeycomb lattice. As illustrated in Fig. \ref{fig:lattice}(a), each spin has a preferred neighbor due to the lattice distortion which picks out pairs of spins to dimerize in the ground state.

Applying a magnetic field to the singlet state generically leads to a BEC transition where a triplet band becomes degenerate with the $S = 0$ ground state, creating an XY antiferromagnet. In typical experiments \cite{RevModPhys.86.563}, it has been found that strengthening this field eventually polarizes the system; no other phase transitions are observed. Recently, experiments on the compound $\text{Yb}_{2}\text{Si}_{2}\text{O}_{7}$ have challenged this paradigm by suggesting the presence of an intermediate magnetic phase with an unknown underlying order \cite{PhysRevLett.123.027201}. This Letter proposes a modification to the Heisenberg model whose ground state order is consistent with all available thermodynamic data and allows for the possibility of such a phase diagram.

On the breathing honeycomb lattice, the Heisenberg model in a magnetic field only realizes the previously mentioned singlet, XY antiferromagnet and polarized phases. In order to model the additional phase observed experimentally, we generalize the Heisenberg model by introducing two forms of anisotropy:

\begin{equation}\label{model}
H = \sum_{\langle ij\rangle,\alpha} J_{ij}^{\alpha}S_{i}^{\alpha}S_{j}^{\alpha} - h\sum_{i,\alpha}g_{z\alpha}S_{i}^{\alpha}
\end{equation}

Here $i,j$ index lattice sites and $\alpha =x,y,z$ are Cartesian indices. The $x,y,z$ directions correspond, repectively, to the $a^{*},b=b^{*},$ and $c$ axes of the C2/m lattice structure. In other words, we are considering an XYZ model for a breathing lattice, and allowing for the possibility that the $z$-axis is not a principal axis of the $g$-tensor. The true ``minimal model'' for the physics of interest is significantly more restricted: it is sufficient to take $J_{ij}^{y}>J_{ij}^{x}=J_{ij}^{z}$ (for all $i,j$) and $g_{zy}=0$, as $y$ is a principal axis. More precise constraints discussed below are imposed by consistency with experiments.

As we will see, there is a regime of parameters which yields the phase diagram in Fig. \ref{fig:lattice}(b). This phase diagram matches thermodynamic data by providing a mechanism for both breaking and restoring an Ising symmetry as an external magnetic field is tuned. For $H_{c_{1}}<H<H_{m}$, the ground state breaks a $\mathbb{Z}_{2}$ symmetry, while for $H>H_{m}$ the system exhibits no symmetry breaking. Importantly, these effects are observable with weak anisotropy: we believe this can explain the coexistence of familiar and unfamiliar features found in Yb$_2$Si$_2$O$_7$ \cite{PhysRevLett.123.027201}.

We will use a variety of complementary techniques to develop a theory which accounts for the observations of Yb$_2$Si$_2$O$_7$. In order to motivate our model (\ref{model}), we begin with a review of salient experimental facts. We then use a linked cluster expansion to compute the triplon spectrum and locate the critical points of the pure Heisenberg model, $H_{c_{1}}$ and $H_{c_{2}}$. Our results are consistent with experimental findings and confirm that the Heisenberg model captures important aspects of the physics of Yb$_2$Si$_2$O$_7$. Spin-wave theory is then applied to the full Hamiltonian (\ref{model}) to show that the perturbations we have introduced produce dispersion relations consistent with neutron scattering data. Having established the consistency of our model with existing data, we develop a qualitative understanding of the new order induced by these perturbations. This is facilitated by a simple self-consistent mean field theory, which reveals the previously undetermined ground state order to be a canted antiferromagnet. This physical picture is then quantitatively verified via a density matrix renormalization group (DMRG) analysis, and our concluding remarks suggest possible experimental tests of our proposals.

\itshape Experimental Considerations.---\normalfont Plausible modifications to the Heisenberg model are strongly constrained by the available experimental data. To establish constraints on the parameters introduced in (\ref{model}), we review the salient details of the experimental work \cite{PhysRevLett.123.027201}.

\begin{enumerate}

\item\itshape Critical fields and zero-field specific heat are modeled well by the pure Heisenberg model\normalfont . In Ref.~\onlinecite{PhysRevLett.123.027201}, it was demonstrated that the Heisenberg model  fits zero-field specific heat data. Below, we will show that the Heisenberg model is consistent with the empirical values of $H_{c_{1}}$ and $H_{c_{2}}$, and they do not depend sensitively on weak perturbations.

\item\itshape The XY antiferromagnet hosts an approximate Goldstone mode\normalfont . Within the energy resolution of the experiment, there is a gapless mode in the band structure of the planar antiferromagnet.

\item\itshape Singularities in the specific heat present in weak fields vanish with increasing field. \normalfont In weak fields, an Ising-like singularity is observed as a function of temperature. Increasing the field to $H_{m}\approx 1.2$ Tesla removes the singularity and leads to smooth behavior as a function of temperature. Further, ultrasound velocity and neutron scattering measurements offer additional evidence for the existence of a phase transition at $H_{m}$.

\end{enumerate}

The first point confirms that the Heisenberg model is a good starting point for an analysis of $\text{Yb}_{2}\text{Si}_{2}\text{O}_{7}$. The second implies that any XY symmetry breaking in the Hamiltonian is bounded from above. The third suggests that the ground state breaks different symmetries as a function of magnetic field. Moreover, the ground state for $H>H_{m}$ smoothly crosses over to the polarized limit when $H= H_{c_{2}}$.

\itshape Phenomenology of the Model.---\normalfont The perturbations to the Heisenberg model which we have introduced are designed to respect these experimental constraints. The key changes are to the XY Heisenberg couplings, $J_{ij}^{y}=\left(1+\lambda\right)J_{ij}^{x}$, and a staggered $g$-tensor component $g_{zx} \ll g_{zz}$, $g_{zx}^{A}=-g_{zx}^{B}$, where $A,B$ are sublattices. By choosing $\lambda\ll 1$, the first two experimental points are addressed: many qualitative features of the Heisenberg model are preserved and the Goldstone mode is only weakly gapped. The staggered $g$-tensor creates a field-dependent competition between antiferromagnetic orders in the X-Y plane. In weak magnetic fields ($H_{c_{1}}<H<H_{m}$), the YY coupling dominates, and the ground state breaks the  $\mathbb{Z}_{2}$ symmetry of the Hamiltonian. In larger magnetic fields ($H>H_{m}$), no symmetry is broken because the $g$-tensor selects a unique antiferromagnetic X-order. Since it breaks no symmetries, this state can cross over smoothly to the polarized limit ($H>H_{c_{2}}$).

\begin{figure}[t!]
  \includegraphics[width=0.99\linewidth]{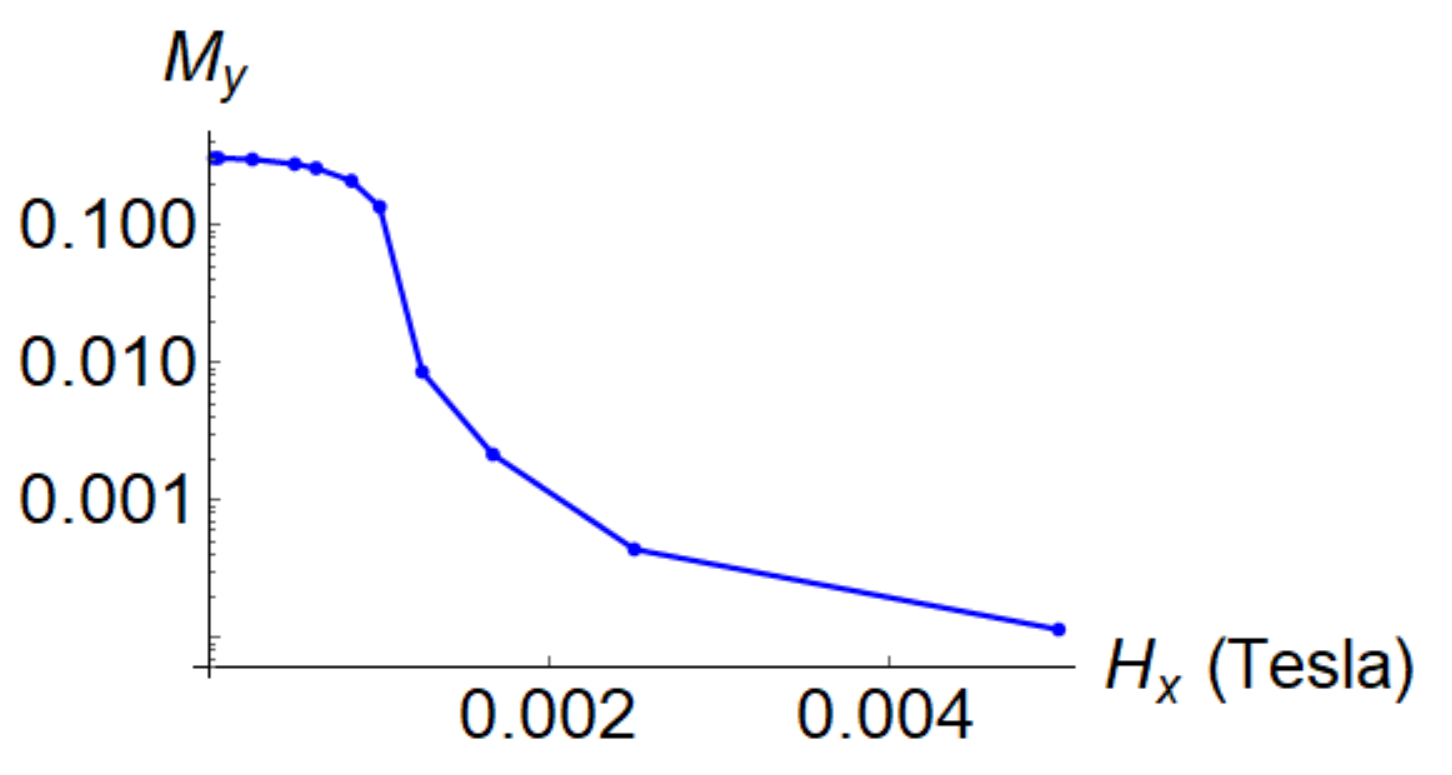}
  \caption{Dependence of $M_y$ on the magnitude of the staggered field $H_x$ ($H_z=0.9$T for each point).  The value of $M_y$ drops off rapidly with $H_{x}$, indicating an instability of the symmetry-breaking mean-field solution to anisotropy in the $g$-tensor. Weakness of the anisotropy is therefore critical to the physics.}
  \label{fig:Myvshx}
\end{figure}

We note that a staggered $g$-tensor is forbidden by the inversion symmetry of the C2/m crystal structure. However, weak deviations from this structure due to lattice distortions are not ruled out by current experimental data. Such a distortion has clear experimental signatures (see the concluding section). The required weakness of our staggered $g$-tensor (see Fig. \ref{fig:Myvshx} and surrounding discussions) is consistent with a distortion-based explanation.

Further, we have explored similar models with uniform $g$-tensors and found that they do not reproduce the phase diagram of Fig. \ref{fig:lattice}. Essentially, a uniform $g$-tensor does not lead to a field-dependent competition between antiferromagnetic orders: the spins have a polarization in the $x-z$ plane proportional to the effective field in each direction. The absence of order in X and Z prevents the formation of the observed phase for $H_{c_{m}}<H<H_{c_{2}}$. This phenomenological description is supported by DMRG data.

The parameters we will choose throughout the paper are $\lambda = 0.03$ and $g_{zx} = g_{zz}/100$. We take the $x$-component of the Heisenberg coupling to be the value obtained experimentally for the isotropic Heisenberg model, $J_{1}^{x} = 0.2173$ meV, $J_{2}^{x} = 0.0891$ meV. Conversions to physical magnetic fields are done with $g$-factors measured in \cite{PhysRevLett.123.027201}. We have found that our results do not qualitatively depend on these choices except in our DMRG analysis, where this issue is discussed.

\itshape Linked-Cluster Expansion.---\normalfont Here we simplify to the isotropic Heisenberg model and assume the $z$-axis is a principal axis of $g$ ($g_{z\alpha} \propto \delta_{z\alpha}$). We will perturbatively compute the critical fields of the BEC transition and show that the result is consistent with experiments. In the limit $J_{2}=h=0$, the ground state of (\ref{model}) is a collection of independent spin singlets. For finite $J_{2}$ with $J_{2}/J_{1} \ll 1$, the ground state remains in the $S=0$ sector with a gap to mobile triplet excitations. These ``single-particle'' states are not eigenstates of (\ref{model}), and to compute their spectrum we use the linked cluster formalism. This allows us to compute the triplet spectrum perturbatively in $J_{2}/J_{1}$ in the thermodynamic limit  \cite{Gelfand1990,2000AdPhy..49...93G,oitmaa_hamer_zheng_2006}.

The spectrum resulting from this analysis has a minimum at $\vec{k} = 0$. We find that (defining $J_{2}/J_{1} = \alpha$)

\begin{equation}\label{bandminimum}
\omega\left(\vec{k}=0\right) = J_{1}\left(1-\alpha-\alpha^{2}+\frac{5}{16}\alpha^{3} + \mathcal{O}\left(\alpha^{4}\right)\right)
\end{equation}

For $h\neq 0$, the $S^{z} = 1$ triplet band decreases linearly in energy leading to a gap closing. The resulting BEC transition has been studied extensively \cite{RevModPhys.86.563,PhysRevLett.86.1082,TANAKA20071343,PhysRevB.37.4936,PhysRevB.75.140403,Nohadani_2004,Giamarchi_2008}. Choosing the couplings and gyromagnetic factors reported in Ref.~\onlinecite{PhysRevLett.123.027201}, we find the critical field $H_{c_{1}} \approx 0.434$ Tesla, in rough agreement with the experimental data. The upper critical field, $H_{c_{2}}$, of the Heisenberg model can be calculated exactly by considering the energetic cost of a spin flip in the polarized phase. We find $H_{c_{2}} = J_{1}+2J_{2} \approx 1.42$ Tesla, also in agreement with experiment.

The singlet ansatz for the ground state is not strictly correct in the presence of anisotropy when $h\neq 0$. However both mean-field and DMRG analyses indicate that the system becomes effectively non-magnetic below $H_{c_{1}}$ in the presence of weak anisotropy (see Fig. \ref{fig:allM}). The agreement between these critical fields and the experimental results justifies our focus on perturbative adjustments to the Heisenberg model.

\begin{figure}
  \includegraphics[width=0.95\linewidth]{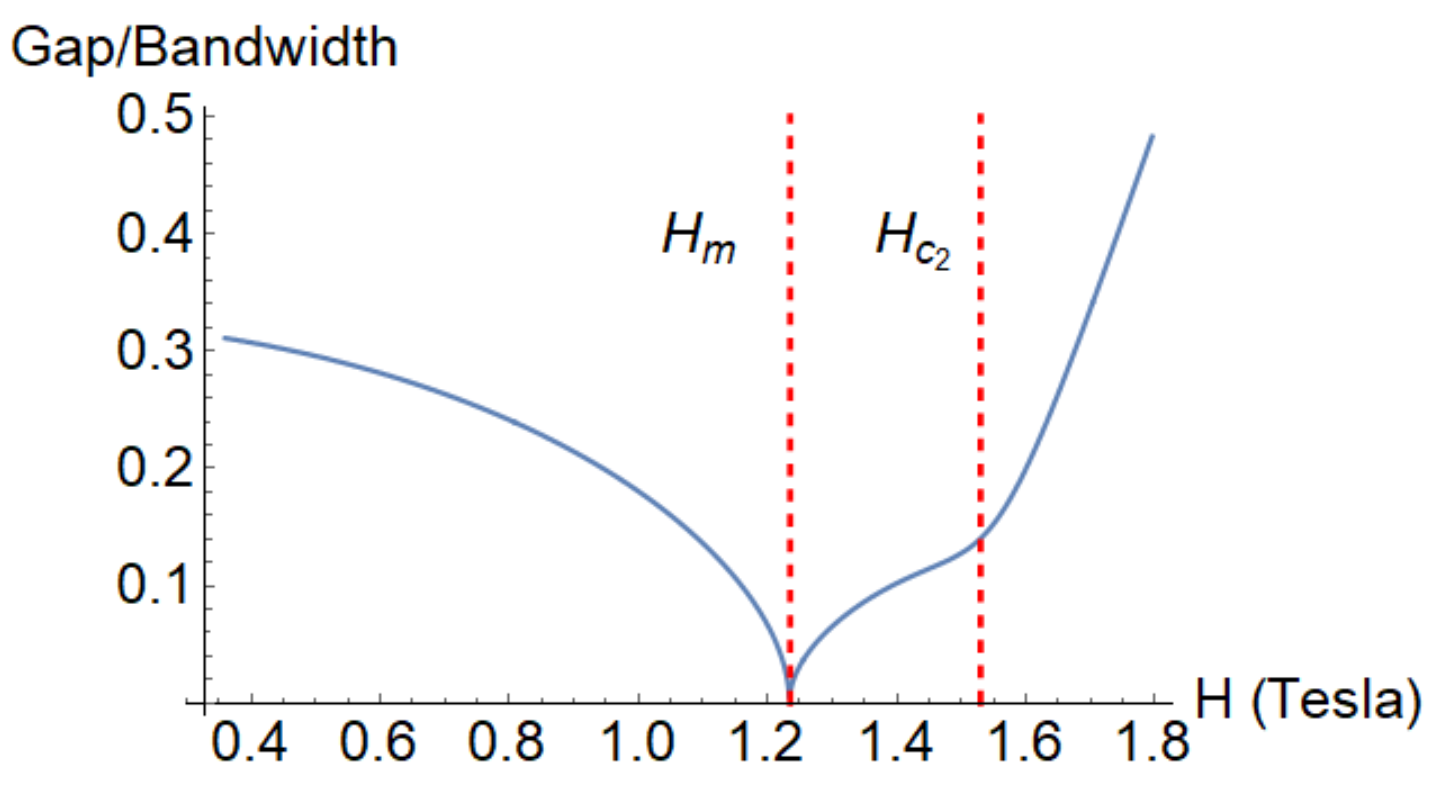}
  \caption{Gap:Bandwidth ratio as a function of field in linear spin-wave theory. Other than the phase transition between spin wave solutions at $H=H_{m}$ (see text), the system is gapped with an energy scale consistent with the energy resolution of available neutron scattering data. For $H>H_{c_{2}}$ the band energy scales linearly.}
  \label{fig:spinwave}
\end{figure}

\itshape Spin-Wave Theory.---\normalfont By introducing anisotropy to the Heisenberg couplings, we have broken the XY symmetry of the model. We therefore anticipate that the spectrum is gapped, and the Goldstone mode observed experimentally is in fact massive. Here we will use linear spin-wave theory to compute the spectrum and show that the anisotropy-induced gap is consistent with the energy resolution of available experimental data.

Our ansatz for the classical spin orientations on sublattices $A,B$ is for a canted antiferromagnet:

\begin{equation}\label{ansatz}
\begin{split}
\vec{S}_{A} = S\left(\sin\theta\cos\phi,\sin\theta\sin\phi,\cos\theta\right) \\ \vec{S}_{B} = S\left(-\sin\theta\cos\phi,-\sin\theta\sin\phi,\cos\theta\right)
\end{split}
\end{equation}

Minimizing the Hamiltonian as a function of $\theta,\phi$ yields two solutions. In weak fields,

\begin{equation}\label{sol1}
\begin{split}
\cos\theta = \frac{h_{z}}{S\left(\bar{J}_{z}+\bar{J}_{y}\right)} \\ \cos\phi = \frac{h_{x}\left(\bar{J}_{x}+\bar{J}_{y}\right)}{\left(\bar{J}_{y}-\bar{J}_{x}\right)\sqrt{S^{2}\left(\bar{J}_{x}+\bar{J}_{y}\right)^{2}-h_{z}^{2}}}
\end{split}
\end{equation}

Here  $\bar{J}_{\alpha} = J_{1}^{\alpha}+2J_{2}^{\alpha}, h_{z} = g_{zz}h, h_{x} = g_{zx}h$. The critical field $H_{m} \approx 1.2$ Tesla is given by $\cos\phi = 1$, and agrees with experimental data. For $H>H_{m}$ the system transitions to

\begin{equation}\label{sol2}
\begin{split}
\phi = 0 \\ \sin\theta = \frac{h_{z}\tan\theta - h_{x}}{S\left(\bar{J}_{z}+\bar{J}_{x}\right)}
\end{split}
\end{equation}

Using the Holstein-Primakoff mapping to bosons, we obtain a quadratic Hamiltonian which can be diagonalized using standard techniques \cite{PhysRev.58.1098,Mourigal_2013,Zhitomirsky_2013}. From the resulting dispersion, we extract the gap and bandwidth; their ratio is shown in Fig. \ref{fig:spinwave}. The bands are gapped everywhere except at $H_{m}$, which separates the spin-wave solutions. The gap:bandwidth ratio is consistent with experimental results, where an energy resolution of $\sim$0.037 meV is available for a band of width $\sim$0.1 meV. We note that the curvature of the excited band also changes as a function of field, in agreement with \cite{PhysRevLett.123.027201}.

\begin{figure}
  \includegraphics[width=0.95\linewidth]{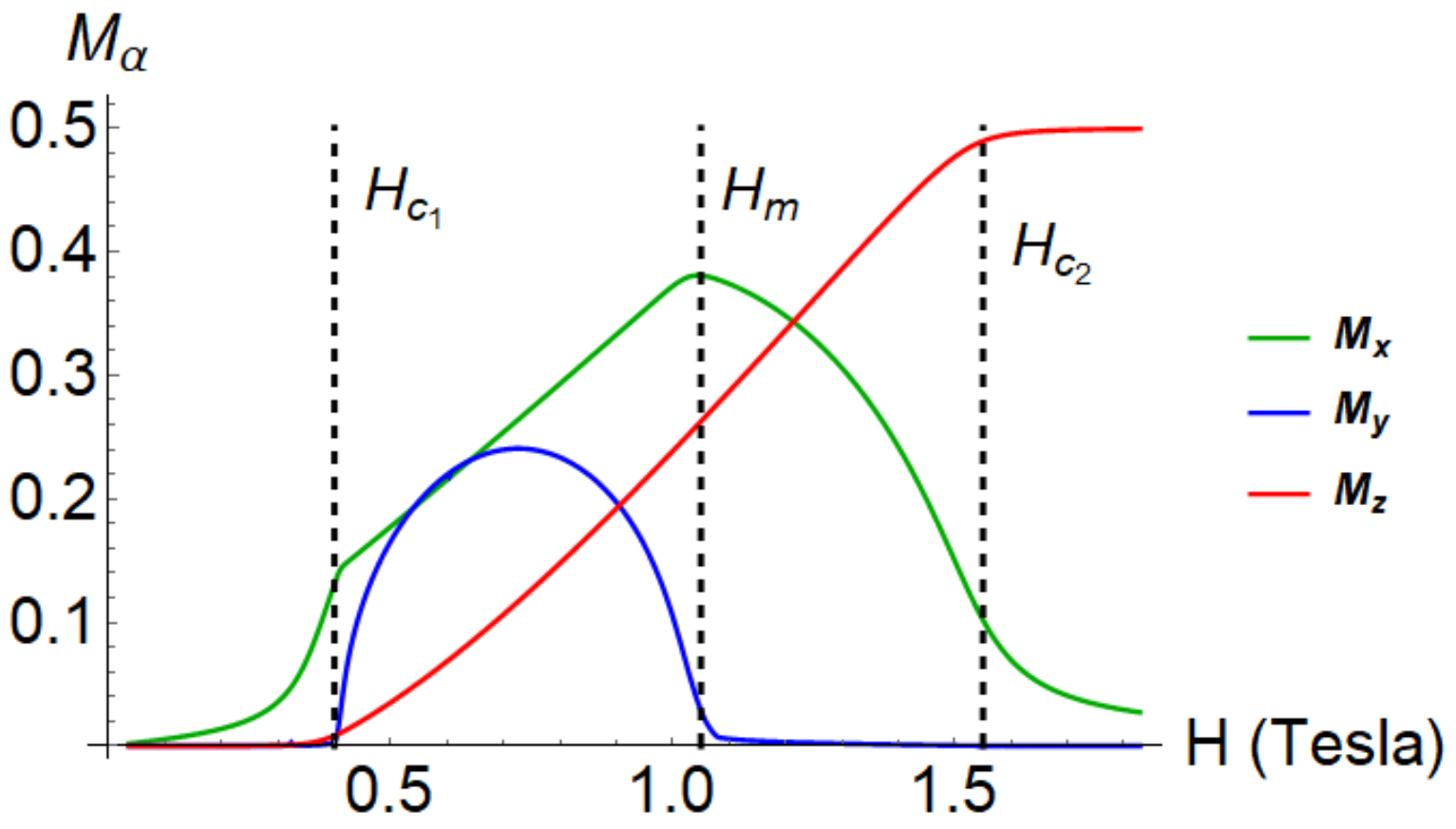}
  \caption{Bulk spin expectation values as a function of magnetic field obtained from self-consistent mean field theory. Note X and Y moments are staggered while Z is uniform. A nonzero $g_{xz}$ changes the weak ($H<H_{c_{1}}$) and strong field ($H>H_{c_{2}}$) phases by forcing $M_{x}\neq 0$. The presence of a nonzero $M_{y}$ for $H_{c_{1}}<H<H_{m}$ indicates $\mathbb{Z}_{2}$ symmetry breaking and corresponds to the standard magnetic phase observed on the high-field side of the BEC phase transition without anisotropy. The range $H_{m}<H<H_{c_{2}}$ corresponds to a canted antiferromagnet which crosses over to the saturated regime at $H_{c_{2}}$.}
  \label{fig:mft}
\end{figure}

\itshape Cluster Mean Field Theory.---\normalfont Given that our model does not contradict the experimental data, we move on to develop a qualitative understanding of the ground states of (\ref{model}). We begin by formulating a mean-field theory using the bipartite structure of the honeycomb lattice. Let $\vec{M}_{A},\vec{M}_{B}$ denote the average magnetizations on sublattices $A, B$. The enhanced coupling $J_{1}$ between neighbors along $y$ suggests that the fundamental degree of freedom is a dimer containing spins $\vec{S}_{A},\vec{S}_{B}$ embedded in an effective magnetic field. The Hamiltonian is

\begin{equation}\label{clustermft}
\begin{split}
H = J_{1}^{\alpha}S_{A}^{\alpha}S_{B}^{\alpha} + 2J_{2}^{\alpha}\left(S_{A}^{\alpha}M_{B}^{\alpha}+S_{B}^{\alpha}M_{A}^{\alpha}\right) \\ - h\sum_{\alpha}\left(g_{z\alpha}^{A}S_{A}^{\alpha}+g_{z\alpha}^{B}S_{B}^{\alpha}\right)
\end{split}
\end{equation}

We take $g_{zy}=0, g_{zx}\ll g_{zz}$. The Hamiltonian (\ref{clustermft}) is analyzed with self-consistent methods, feeding in an ansatz for $\vec{M}_{A},\vec{M}_{B}$ and calculating new values $\vec{M}_{i}\equiv \langle\psi |\vec{S}_{i}|\psi\rangle $, where $|\psi\rangle$ is the instantaneous ground state. These values are updated until convergence is achieved.

For sufficiently small $g_{zx}$, we find that the solution in Fig. \ref{fig:mft} is the most energetically favored. For small fields ($H<H_{c_{1}}$), the solution is only weakly magnetic due to the staggered field induced by $g_{zx}$. Between the critical fields $H_{c_{1}}<H<H_{c_{2}}$, two phases appear, distinguished by the value of $M_{y}$. The first ($H<H_{m}$) exhibits $\mathbb{Z}_{2}$ symmetry breaking and accounts for the singularity observed in the specific heat; the latter exhibits the high-field crossover behavior required by the absence of thermodynamic singularities. This previously unidentified phase is a canted $XZ$ antiferromagnet.

We note the existence of another mean-field solution in which $M_{y}=0$ everywhere. This case does not support the experimental data as it has no symmetry breaking. The energetic favorability of one solution over another depends on the precise anisotropy parameters chosen; it is unclear how quantum fluctuations will impact that selection. Further, it is not obvious that the inter-dimer coupling $J_{2}$ is sufficiently small to justify a mean-field description. To address these concerns, we employ DMRG to investigate the stability of our results. There we find that both mean field solutions survive quantum fluctuations and remain energetically competitive. Further, there is a regime of parameters in which the solution in Fig. \ref{fig:mft} is favored.

\begin{figure}
  \includegraphics[width=0.95\linewidth]{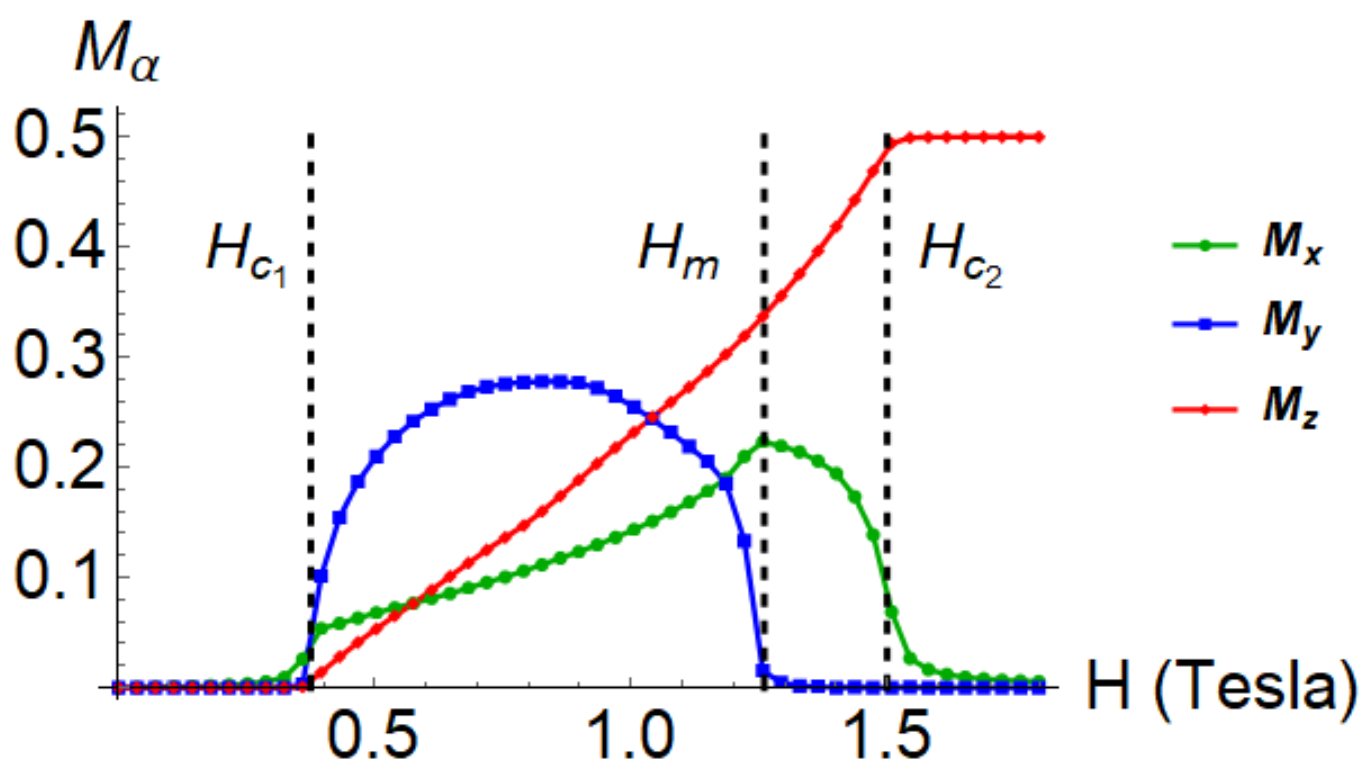}
  \caption{All components of magnetization obtained from DMRG. Note X and Y moments are staggered while Z is uniform. For fields in the range $[0.35,1.25]$T, a magnetic phase with $\mathbb{Z}_{2}$ symmetry breaking is observed. In the range $[1.25,1.6]$T we find another phase with $M_{y}=0$ and a large $M_{x}$. This phase crosses over smoothly to the polarized limit. The qualitative agreement with the mean-field phase diagram of Fig. \ref{fig:mft} confirms the accuracy of that analysis. Here $g_{xz}=g_{zz}/500$ is chosen.}
  \label{fig:allM}
\end{figure}

\itshape DMRG Analysis.---\normalfont To verify the mean-field solution, we use DMRG to compute ground state expectation values \cite{white1992density}. This tensor network method efficiently simulates systems which are well-described by the matrix product state (MPS) ansatz \cite{baker2019m,affleck1987rigorous,verstraete2006matrix,schollwock2005density,schollwock2011density}. Our system is studied on a cylinder with a width of four dimers and 128 total spins.

We use a single-site representation of the renormalized tensor network to update each step \cite{hubig2015strictly} with the Hamiltonian (\ref{model}). To guarantee that the proper symmetry sector is obtained, we apply pinning fields on the open boundaries of the system to break the $\mathbb{Z}_{2}$ symmetry of the Hamiltonian. The pinning field is removed after two DMRG sweeps, and we find that in the symmetry breaking region this produces a lower-energy state than unbiased DMRG.

From the resulting ground-state wavefunction, local measurements of quantities $M_\alpha=\sqrt{\sum_{i=1}^{N_s}\langle \hat S_i^\alpha\rangle^2}/N_{s}$ are performed.  The results are shown in Fig.~\ref{fig:allM} and closely match those from mean-field theory.  A dome in $M_y$ appears in the regime  $H_z\in[0.35,1.25]$T which indicates symmetry breaking. The sizable $M_{x}$ values for $H_{z}\in [1.25,1.6]$T differentiate that region from the saturated limit. In this case, we see that the mean field solutions are sufficient to capture the essential physics of the system, albeit with different parameters in the Hamiltonian.

The results in Fig. \ref{fig:allM} are found with $g_{xz}=g_{zz}/500$.  This value is arbitrary and can affect which mean field solution is obtained; to account for this, Fig. \ref{fig:Myvshx} shows the dependence of the symmetry-breaking order parameter $M_{y}$ on the staggered field $H_x$ with fixed (uniform field) $H_z$. The solutions were found by first tuning $H_z$ to $0.9$T with pinning fields. The pinning fields are then removed and the staggered field is increased.  The ground state changes from a Y-ordered antiferromagnet to a state where $M_y = 0$ as $H_x$ increases. The instability of the symmetry-breaking solution to anisotropy in the $g$-tensor reveals that $g_{zx}$ is necessarily small. This is consistent with $g_{zx}\neq 0$, which requires deviations from the C2/m crystal structure currently proposed experimentally. In principal, such deviations can be observed directly with higher-resolution scattering data. Importantly, the qualitative features of the phase diagram should be robust to other perturbations.

\itshape Conclusions.---\normalfont With a variety of theoretical techniques, we have constructed an explanation for the experimentally proposed phase diagram of Yb$_{2}$Si$_{2}$O$_{7}$. These techniques complement each other; each of them supports the physical picture presented in this Letter. We emphasize again that \itshape small \normalfont perturbations to the Heisenberg model can explain the observed thermodynamic responses of the material.

Experimental verification of these details remains crucial, and our theory suggests natural tests of itself. The structure of local magnetic moments in the material can be probed with nuclear magnetic resonance (NMR) techniques. In particular, we anticipate planar antiferromagnetic order which collapses onto one axis in the plane with increasing field. Particularly important are measurements of X-moments in the material: a staggered magnetization in that case would confirm that a C2/m forbidden, staggered $g$-tensor is crucial to describing Yb$_2$Si$_2$O$_7$. Further, more precise neutron scattering measurements may reveal a spin gap for $H_{c_{1}}<H<H_{m}$, the magnitude of which will constrain the necessary perturbations of the Heisenberg model.

\itshape Acknowledgements.---\normalfont We thank Leon Balents and Fr\'ed\'eric Mila for suggestions in the development of the mean-field theory. We are also grateful to Miles Stoudenmire for helpful comments on DMRG techniques.

We thank the Institute for Complex Adaptive Matter (ICAM)-supported  school on Emergent Phenomena in Correlated Quantum Matter in Carg\'ese, where this collaboration was initiated.

The work of M. Flynn and R.R.P. Singh is supported in part by NSF DMR grant number 1855111. The work of S. Jindal is supported in part by NSF PHY grant number 1852581. T.E.B.~thanks the support of the Postdoctoral Fellowship from Institut quantique and support from Institut Transdisciplinaire d'Information Quantique (INTRIQ).  This research was undertaken thanks in part to funding from the Canada First Research Excellence Fund (CFREF).  This research was enabled in part by support provided by Calcul Qu\'ebec (www.calculquebec.ca) and Compute Canada (www.computecanada.ca).  Computations were made on the supercomputer Mammouth (mp2), located at Universit\'e de Sherbrooke.

\bibliography{QuantumDimerModel}{}

\end{document}